\documentclass[9pt,a4paper,twocolumn]{scrartcl}

\usepackage[TS1,T1]{fontenc}
\usepackage[utf8]{inputenc}
\usepackage{lmodern}
\usepackage{xspace}
\usepackage[dvipdfmx]{hyperref}
\hypersetup{bookmarksnumbered=true,
bookmarks,
hypertexnames=false,
pdfauthor={Serge Abiteboul, Émilien Antoine, Julia Stoyanovich},
pdftitle={webdam system}, pdfsubject={webdamlog implementation},
pdfkeywords={webdamlog, datalog, distributed, declarative, provenance,graph},
pdfcreator={latex with hyperref pa
ckage},
pdfproducer={dvipdfmx}}
\usepackage[hyperpageref]{backref}
\usepackage{graphicx}
\DeclareGraphicsExtensions{.eps}
\usepackage{color}
\usepackage{multirow}
\usepackage{balance}

\definecolor{darkgreen}{rgb}{0,0.5,0}

\newcommand{\serge}[1]{[\textcolor{red}{{\bf Serge: }#1}]}

\newcommand{\dollar}{{\$}}

\newcommand{\wdm}{\emph{Webdam}\xspace}
\newcommand{\wdl}{\emph{WebdamLog}\xspace}
\newcommand{\wde}{\emph{WebdamLog system}\xspace}
\newcommand{\bud}{\emph{Bud}\xspace}

\newcommand{\peer}[1]{\ensuremath{\mathsf{#1}}}
\newcommand{\relName}[1]{\ensuremath{\mathsf{#1}}}
\newcommand{\relduo}[2]{\relName{#1}\textbf{@}\peer{#2}}
\newcommand{\reltri}[3]{\relName{#1}\textbf{@}\peer{#2}(\ensuremath{\mathsf{#3}})}
\newcommand{\var}[1]{\ensuremath{\mathsf{\$}\mathsf{#1}}}
\newcommand{\tuple}[1]{\ensuremath{\overline{\mathsf{#1}}}}
\newcommand\eat[1]{}

\newcounter{rulecounter} 
\newcounter{subrulecounter}[rulecounter] 

\title{The Webdamlog System}

\subtitle{Managing Distributed Knowledge on the
Web
\footnote{This work has been partially funded by the European
Research Council under the European Community's Seventh Framework
Programme (FP7/2007-2013); ERC grantWebdam, agreement
226513. \url{http://webdam.inria.fr/}} }

\author{
Serge Abiteboul\\
Inria Saclay \& ENS Cachan\\
France\\
first.last@inria.fr
\and
Émilien Antoine\\
Inria Saclay \& ENS Cachan\\
France\\
first.last@inria.fr
\and
Julia Stoyanovich\\
Drexel University, USA\\
Skoltech, Russia\\
stoyanovich@drexel.edu
}

\begin{document}

\maketitle

\begin{abstract}
  We study the use of \wdl, a declarative high-level language in the
  style of datalog, to support the distribution of both {\em data} and
  {\em knowledge} (i.e., programs) over a network of autonomous
  peers. The main novelty of \wdl compared to datalog is its use of
  delegation, that is, the ability for a peer to communicate a program
  to another peer.

  We present results of a user study, showing that users can write
  \wdl programs quickly and correctly, and with a minimal amount of
  training.  We present an implementation of the \wdl inference engine
  relying on the \bud datalog engine.  \eat{A key issue in this
    setting is the rapid reaction to changes in data and knowledge. We
    show how to efficiently support such changes using a novel form of
    provenance graph.  We describe an extensive experimental
    evaluation of the \wdl engine, demonstrating that \wdl can be
    implemented efficiently.} We describe an experimental evaluation
  of the \wdl engine, demonstrating that \wdl can be implemented
  efficiently.  We conclude with a discussion of ongoing and future
  work.

\end{abstract}


\section{Introduction}
\label{sec:intro}


A number of works have argued for developing a holistic approach to
distributed content management, e.g. \emph{P2P Content
  Warehouse}~\cite{AbiteboulP2PCW},
\emph{Dataspaces}~\cite{FranklinHM05} and \emph{Data
  rings}~\cite{AbiteboulP07}.  The goal is to facilitate the
collaboration of autonomous peers towards solving content management
tasks. Such situations arise for instance in personal information
management (PIM), which is often given as an important motivating
example~\cite{FranklinHM05}.  In~\cite{AbiteboulP07}, the authors
argued for founding such data exchange on declarative languages, to
facilitate the design of applications, notably by non-technical users.

In the present work, we propose an approach for managing data and
knowledge (i.e., programs) over a network of autonomous peers. From a
system viewpoint, the different actors are autonomous and
heterogeneous in the style of P2P~\cite{AbiteboulP07,
  FranklinHM05}. However, we do not see the system we developed as an
alternative to existing network services such as Facebook or
Flickr. Rather, we view our system as the means of seamlessly managing
distributed knowledge residing in any of these services, as well as in
a wide variety of systems managing personal or social data.

Our system uses the \wdl language~\cite{webdamlog}, a declarative
high-level language in the style of datalog, to support the
distribution of both {\em data} and {\em knowledge} (i.e., programs)
over a network of autonomous peers.  In recent years, there has been
renewed interest in using languages in the datalog family in a large
range of applications, from program analysis, to security and privacy
protocols, to natural language processing, to multi-player games.  The
arguments in favor of datalog-style languages are familiar ones: a
declarative approach alleviates the conceptual complexity on the user,
while at the same time allowing for powerful performance optimizations
on the part of the system.

\wdl is a datalog-style language that emphasizes cooperation between
autonomous peers communicating in an asynchronous manner.\eat{ We
  briefly recall \wdl (Section~\ref{sec:wdl}).} The \wdl language
extends datalog in a number of ways, supporting
updates\eat{~\cite{AVDatalogUpdate,dedalus}}~\cite{dedalus},
distribution~\cite{activexml-vldb}, negation~\cite{well-founded}, and,
importantly, a novel feature called
delegation~\cite{webdamlog}. As a result, \wdl is neither as
simple nor as beautiful as datalog. It is also more procedural, which
is needed to capture real Web applications with the peers' knowledge
evolving over time.

We illustrate by example (Section~\ref{sec:exa}) that the language
(formally recalled in Section~\ref{sec:wdl}) is indeed well adapted to
specifying realistic distributed content management tasks, notably in
PIM.  Our technical contributions are described in the following
sections:

\begin{itemize}

\item We present results of a user study, showing that users can write
  \wdl programs quickly and correctly, and with a minimal amount of
  training (Section~\ref{sec:userstudy}).

\item We present an implementation of the \wdl engine relying on the
  \bud datalog engine (Section~\ref{sec:system}).  Our implementation
  supports novel linguistic features such as peer and predicate
  variables and rule delegation.

\eat{
\item A key issue in this setting is the rapid reaction to changes in
  data and knowledge. We show how to efficiently support such changes
  using a novel form of provenance graph (Section~\ref{sec:dynamic}).}

\item We describe an experimental evaluation of the \wdl engine
  (Section~\ref{sec:experiments}).

\end{itemize}

We discuss related work in Section~\ref{sec:related}, outline
future research directions and conclude in Section~\ref{sec:conc}.

\section{Running example}
\label{sec:exa}

Suppose that Alice and Bob are getting married, and their friends want
to offer them an album of photos in which the bride and groom appear
together.  Such photos may be owned by friends and family members of
Alice and Bob.  Owners of the photos may store them on a variety of
services and devices, including, e.g., desktop computers, smartphones,
Picasa, and Flickr.

Making a photo album for Alice and Bob involves the following steps:
(1) Identify friends of Alice and Bob using Facebook and Google+; (2)
Find out where each friend keeps his/her photos and how to access
them; (3) From among all photos that are obtained, select those that
feature both Alice and Bob, using, e.g., tags or face recognition
software; and (4) Ask Sue, a friend of Alice, to verify that the
selected photos are appropriate for the photo album and to possibly
exclude some from this album.

As should be clear from the example, such a task would be much more
manageable if it were executed automatically.  Its execution involves
a certain amount of simple reasoning on the part of the system, which
can be naturally specified with declarative rules.  For example, for
Step (1), the following \wdl rule computes the union of Alice's and
Bob's Facebook contacts in a relation \relName{allFriends} on Sue's
peer\eat{ (that we call Sue to simplify)}:
\begin{verbatim}
[rule at sue]
allFriends@sue($name) :- friends@aliceFB($name)
allFriends@sue($name) :- friends@bobFB($name)
\end{verbatim}
using wrappers to Facebook for Alice and Bob.

In general, a \emph{peer name} such as \peer{aliceFB} or
\peer{sueIPhone} denotes a system or a device associated to a
particular URL. Also a \emph{relation name} such as \relName{friends}
or \relName{contacts} denotes the name of a relation or a service on
the corresponding system/device.

For simplicity, we assume that a person's name, e.g. \peer{alice},
corresponds to the name of the {\em peer} that the particular friend
uses as entry point to the \wdm system. (This name is thus
associated to a particular URL.)  We assume that each such peer keeps
localization data for the corresponding person. For instance, relation
\relName{photoLocation} in that peer tells where (i.e., at which
peers) this person keeps her photos.  The following rule, at peer
\peer{sue}, {\em delegates} Steps (2) and (3) of the photo album task
to the peers corresponding to the peers corresponding to her friends:
\begin{verbatim}
[rule at sue]
album@sue($photo,$name) :-
   allFriends@sue($name),
   photoLocation@$name($peer),
   photos@$peer($photo),
   features@$peer($photo,alice),
   features@$peer($photo,bob)
\end{verbatim}
The key feature of this rule is the use of the \wdl language to
share the work. Let Dan be a friend, and so a possible source.
Then Sue's peer will delegate the following rule to Dan's peer:
\begin{verbatim}
[rule at dan]
album@sue($photo,dan) :-
   photoLocation@dan($peer),
   photos@$peer($photo),
   features@$peer($photo,alice),
   features@$peer($photo,bob)
\end{verbatim}
Now suppose that Dan uses both Picasa and Flickr.  Then, Dan's peer
will delegate to \peer{danPicasa} (a wrapper for Dan's account on
Picasa) the following rule:
\begin{verbatim}
[rule at danPicasa]
album@sue($photo,dan) :-
   photos@danPicasa($photo),
   features@danPicasa($photo,alice),
   features@danPicasa($photo,bob)
\end{verbatim}
and similarly for Flickr.

Note how the tasks are automatically shared by many peers. Observe
that when new friends of Alice or Bob are discovered (e.g., proposed
by some known friends), Sue's album, which is defined {\em
  intentionally}, is automatically updated.  Observe also that, to
simplify, we assume here that all peers use a similar organization
(ontology). This constraint may easily be removed at the cost of
slightly more complicated rules.

Now consider Step (4) in the photo album task. Sue may decide, for
instance, that photos of the couple from Dave's Flickr stream are
inappropriate, and that Dave should be excluded from the set of
sources.  Such manual curation by Sue may be accomplished by modifying
the definition of allFriends:
\pagebreak
\begin{verbatim}
[rule at sue]
allFriends@sue($name) :- friends@aliceFB($name),
                  not blocked@sue($name)
allFriends@sue($name) :- friends@bobFB($name)
                  not blocked@sue($name)
\end{verbatim}
By inserting/removing facts in \relduo{blocked}{sue}, Sue now controls
who can participate. A similar control can also be added at the photo
or photo location level.

Observe that updates result in modifying the programs running at the
participating peers.  For instance, the sets of rules at the various
peers evolve, controlled by Sue's updates as well as by the discovery
of new friends of Alice or Bob, and of new sources of
photos. Consequently, the album evolves as well.

We will use the example of this section throughout the paper to
demonstrate the salient features of our approach.


\section{The WebdamLog Language}
\label{sec:wdl}

In this section, we briefly recall the language \wdl
from~\cite{webdamlog}.

We assume the existence of a countable set of variables and of a
countable set of data values that includes a set of relation names and
a set of peer names. (Relation and peer names are part of the
data.) Variables start with the symbol {\dollar}, e.g. {\dollar}$x$.

{\bf Schema.} A \emph{relation} in our context is an expression $m@p$
where $m$ is a relation name and $p$ a peer name.  A \emph{schema} is
an expression $(\pi, E, I, \sigma)$ where $\pi$ is a possibly infinite
set of peer names, $E$ is a set of extensional relations of the form
$m@p$ for $p \in \pi$, $I$ is a set of intentional relations of the
form $m@p$ for $p \in \pi$, and $\sigma$, the sorting function,
specifies for each relation $m@p$, an integer $\sigma(m@p)$ that is
its sort. A relation cannot be at the same time intentional and
extensional.

{\bf Facts.} A \emph{fact} is an expression of the form $m@p(a_1, ...,
a_n)$, where $n = \sigma(m@p)$ and $a_1, ..., a_n$ are data values.
An example of a fact is:

\reltri{pictures}{myalbum}{1771.jpg, "``Timbuktu''", 11/11/2011}

{\bf Rules.} A \emph{term} is a constant or a variable.  A \emph{rule}
in a peer $p$ is an expression of the form:
\begin{tabbing}
    {[}at~p]~
    \reltri{\var{R}}{\var{P}}{\var{U}}\mbox{:-}\=
    ($\neg$) \reltri{\var{R_1}}{\var{P_1}}{\var{U_1}},\dots,\\
    \> ($\neg$) \reltri{\var{R_n}}{\var{P_n}}{\var{U_n}}\\
\end{tabbing}
where \var{R}, \var{R_i} are relation terms, \var{P}, \var{P_i} are
peer terms, \var{U}, \var{U_i} are vectors of terms.  The following
safety condition is imposed: that \var{R} and \var{P} must appear
positively bound in the body and each variable occurring in a
negative literal must also appear positively bound in the body. In
addition, rules are required from left to right and it is also
required that each peer name \var{P_i} must be positively bound in a
previous atom.

{\bf Semantics.} At a particular point in time, each peer \peer{p} has
a \emph{state} consisting of some facts, some rules specified locally,
and possibly of some rules that have been delegated to $p$ by other
peers.  Peers evolve by updating their base of facts, by sending facts
to other peers, and by updating their delegations to other peers. So,
both the set of facts and the set of delegated rules evolve over
time. (To simplify, we follow \cite{webdamlog} in assuming that
the set of rules specified locally is fixed.)

The semantics of a rule with head \reltri{m}{p}{u} in a peer \peer{p'}
depends on the nature of the relation in its head: whether it is
extensional (\relduo{m}{p} in $E$) or intentional (\relduo{m}{p} in
$I$), and whether it is local (\peer{p}=\peer{p'}) or not.  We first
consider rules in which all relations occurring in the body are local;
we call such rules {\em local rules}. A subtlety lies in the use of
variables for peer names. The nature of a rule may depend on the
instantiation of its variables, i.e., one instantiation of a
particular rule may be local, whereas another may not be.

We distinguish 5 cases identified by a letter in which we classify the
rules.

\medskip\textbf{A. Local rule with local intentional head} (datalog) These rules
define local intentional predicates, as in classic datalog.

\medskip\textbf{B. Local rule with local extensional head} (local database updates)
Facts derived by this kind of rules are inserted into the local
database.  Note that, by default, like in Dedalus\cite{dedalus}, facts are not
persistent. To have them persist, we use rules of the form
\reltri{m}{p}{U} \mbox{:-} \reltri{m}{p}{U}. Deletion can be captured
by controlling the persistence of facts.

The two previous kinds of rules, containing only predicates of the local
peer, do not require network communication, and are not affected by
problems due to asynchronicity of the network.

\medskip\textbf{C. Local rule with non-local extensional head} (messaging) Facts
derived by rules of this kind are sent to other peers. For example,
the rule:
\begin{tabbing}
    {[}at mi] \=\reltri{\var{m}}{\var{p}}{\var{name}, "``Happy~birthday!''"} \mbox{:-}  \\
    \> \reltri{today}{mi}{\var{date}}, \\
    \> \reltri{birthday}{mi}{\var{name},\var{m}, \var{p},\var{date}}
\end{tabbing}
where \peer{mi} stands for my iPhone, results in sending a Happy Birthday
message to a contact on the day of his birthday. Observe that the name
{\dollar}p of the peer and the name {\dollar}m of the message varies
depending on the person.

\medskip\textbf{D. Local rule with non-local intentional head} (view delegation)
Such a rule results in installing a view remotely. For instance, the rule
\begin{tabbing} {[}at mi]
    \=\reltri{boyMeetsGirl}{gossipsite}{\var{girl}, \var{boy}}
    \mbox{:-}  \\
    \> \reltri{girls}{mi}{\var{girl}, \var{loc}},\\
    \> \reltri{boys}{mi}{\var{boy}, \var{loc}}
\end{tabbing}
installs a join of two \peer{mi} relations at \peer{gossipsite}.

Finally we consider non-local rules.

\medskip\textbf{E. Non-local} (general delegation) Consider the rule
\begin{tabbing}
    {[}at mi] \=\reltri{boyMeetsGirl}{gossipsite}{\var{girl}, \var{boy}} \mbox{:-} \\
    \> \reltri{girls}{mi}{\var{girl}, \var{loc}}, \reltri{boys}{ai}{\var{boy},\var{loc}}
\end{tabbing}
where \peer{ai} stands for Alice's iPhone.  This results in installing, at
\peer{gossipsite}, a view \relName{t_{r@mi}} and a rule, defined as follows:
\begin{tabbing}
{[}at \peer{mi}] \=\reltri{t_{r@mi}}{ai}{\var{girl}, \var{loc}} \mbox{:-} \\
\> \reltri{girls}{mi}{\var{girl}, \var{loc}} \\
{[}at \peer{ai}] \reltri{boyMeetsGirl}{gossipsite}{\var{girl}, \var{boy}} \mbox{:-} \\
\> \reltri{t_{r@mi}}{ai}{\var{girl}, \var{loc}}, \reltri{boys}{ai}{\var{boy}, \var{loc}}
\end{tabbing}
Note that both rules are now local. Note also that, when
\relduo{girls}{mi} changes, this modifies the view at Alice's iPhone,
possibly changing the semantics of \relduo{boyMeetsGirl}{gossipsite}.

In~\cite{webdamlog}, we formally define the semantics of \wdl.  We
show that, unless all peers and programs are known in advance,
delegation strictly increases the expressive power of the model.  If
they are known in advance, delegation does not bring any extra
power. Of course, delegation is also useful in practice, because it
enables obtaining logic (rules) from other sites, and deploying logic
(rules) to other sites.  Conditions for systems to be deterministic
are shown in \cite{webdamlog}, and are extremely restrictive. Even in
the absence of negation, a \wdl system will typically not be
deterministic because of asynchronicity.


\section{Usability of WebdamLog}
\label{sec:userstudy}

We argued in the introduction that \wdl can be used to declaratively
specify distributed tasks in a variety of applications, including
personal data management.  We conducted a user study to demonstrate
the usability of \wdl in this particular domain.

{\bf Participants.} We recruited 27 participants for the user
study\eat{ in the US and in France}.  We present a break-down of
results by two groups.

{\em Group 1} consisted of 16 participants with training in Computer
Science.
Among them, 5 had basic database background, and 4 were familiar with
advanced database concepts, including datalog. The group had the
following break-down by highest completed education level: 2 high
school, 3 BS, 9 MS, and 2 PhD.

{\em Group 2} consisted of 11 participants with no CS training, and
with the following break-down by highest completed education level: 3
vocational school, 6 BS, 2 MS.

{\bf Study design.}  All participants were given a brief tutorial in
which basic features of \wdl were explained informally, and
demonstrated through examples.  The tutorial took 15-20 minutes for
{\em Group 1} and 25 minutes for {\em Group 2}.  Following the
tutorial, all participants were asked to take a written test.  The
test consisted of three problems that tested comprehension of
different features of \wdl, including local and non-local rules, rules
with variable relation and peer names, and delegation.  In the
tutorial and the test, we did not make an explicit distinction between
intentional and extensional relations, and we ignored recursion. 


\eat{We now describe the contents of the user study test, which are
  reproduced literally, apart from formatting.}

The user study test had the following contents, reproduced here
literally, apart from formatting.

\textbf{Problem 1.} Consider the following relations and facts.

\begin{small}
\begin{verbatim}
schema: songs(fileName,content) // the same at all peers

songs@lastFM("song1.mp3", "...")
songs@lastFM("song2.mp3", "...")
songs@lastFM("song3.mp3", "...")
songs@pandora("song4.mp3", "...")
songs@pandora("song5.mp3", "...")
\end{verbatim}
\end{small}

\eat{Assume that {\tt songs} relations at all peers have the same schema.}

\begin{enumerate}
\item Write one or several rules that copy all songs from {\tt lastFM}
  and {\tt Pandora} into relation {\tt songs} at peer {\tt myLaptop}.

\item Suppose now that relation {\tt peers}@{\tt myLaptop} contains
  names of peers on which to look for music. You can assume that each
  peer stores songs in a relation called {\tt songs}, with the same
  schema as above.  Write a \wdl program that copies songs from
  all peers into {\tt songs}@{\tt myLaptop}.

\item Write a rule that copies songs from {\tt songs}@{\tt myLaptop}
  into the {\tt songs} relation on each peer whose name is listed in
  {\tt peers}@{\tt myLaptop}.
\end{enumerate}

\textbf{Problem 2.} Consider the following relations and facts.

\begin{small}
\begin{verbatim}
schema: friends(friendName)  photos(fileName,content)
         inPhoto(fileName, friendName)

friends@facebook("ann")
friends@facebook("sue")
friends@facebook("zoe")

photos@ann("sunset.jpg", "...")
photos@ann("vacation.jpg", "...")
photos@ann("party.jpg", "...")

photos@sue("image1.jpg","...")
photos@sue("image2.jpg","...")

inPhoto@ann("vacation.jpg", "jane")
inPhoto@ann("vacation.jpg", "ann")
inPhoto@ann("party.jpg", "jane")
inPhoto@ann("party.jpg", "zoe")
inPhoto@ann("party.jpg", "sue")

inPhoto@sue("image2.jpg", "sue")
inPhoto@sue("image2.jpg", "jane")
\end{verbatim}
\end{small}

\eat{
\begin{small}
\begin{verbatim}
schema: friends(friendName)  photos(fileName,content)
         inPhoto(fileName, friendName)

friends@facebook("ann")
friends@facebook("sue")
friends@facebook("zoe")
                                   inPhoto@ann("vacation.jpg", "jane")
photos@ann("sunset.jpg", "...")    inPhoto@ann("vacation.jpg", "ann")
photos@ann("vacation.jpg", "...")  inPhoto@ann("party.jpg", "sue")
photos@ann("party.jpg", "...")     inPhoto@ann("party.jpg", "jane")
                                   inPhoto@ann("party.jpg", "zoe")

photos@sue("image1.jpg","...")     inPhoto@sue("image2.jpg", "sue")
photos@sue("image2.jpg","...")     inPhoto@sue("image2.jpg", "jane")
\end{verbatim}
\end{small}
}

Assume that {\tt photos} and {\tt inPhoto} relations at all peers have
the same schema.  Consider now the following \wdl rule.

\begin{small}
\begin{verbatim}
photos@myLaptop($X,$Z) :- friends@facebook($Y),
           photos@$Y($X,$Z), inPhoto@$Y($X,"jane")
\end{verbatim}
\end{small}

\begin{enumerate}

\item Explain in words what this rule computes.

\item List the facts in that are in {\tt photos}@{\tt myLaptop} after
  the rule above is executed.

\item List the facts that are in {\tt photos}@{\tt myLaptop} if the
  following rule is executed instead:

\begin{small}
\begin{verbatim}
photos@myLaptop($X,$Z) :- friends@facebook($Y),
           photos@$Y($X,$Z), inPhoto@$Y($X,"jane"),
           inPhoto@$Y($X,"sue")
\end{verbatim}
\end{small}

\end{enumerate}

\textbf{Problem 3.} Recall the example from the tutorial, in which we
looked at subscribing the peer {\tt myLaptop} to CNN news.  This
example is reproduced below.

\begin{small}
\begin{verbatim}
schema: news@cnn(text)    news@myLaptop(source, text)
         subscribers@cnn(peer)

news@cnn("US Olympic gold")
news@cnn("Higgs boson seen in action")
subscribers@cnn("myLaptop")

[at cnn] news@$X("cnn", $Y) :- subscribers@cnn($X),
                               news@cnn($Y)
\end{verbatim}
\end{small}

Suppose that you would now like to receive CNN news on peer {\tt
  myPhone}, and to store them in relation {\tt news}, with the schema
{\tt source,text}. Describe at least 1 method for doing this.  You may
assume that you can add rules at peers {\tt cnn}, {\tt myLaptop} and
{\tt myPhone}, and that you can insert facts into relations on any of
these peers.


{\bf Results.} The results of the study were very encouraging.

{\em Group 1.} On Problem 1, 3 participants received a score of 2.5
out of 3, while 13 participants received a perfect score. All
participants received a perfect score on Problem 2. Problem 3 was
open-ended, and all participants gave at least one correct answer.  4
participants gave 3 correct answers, 4 gave 2 correct answers (2 of
these also gave 1 incorrect answer each), and the remaining 8
participants each gave 1 correct answer.

We also asked participants to record how long it took them to answer
each problem, in minutes.  Problem 1 took between 2.5 and 6 minutes,
Problem 2 between 2 and 9 minutes, and Problem 3 between 1 and 8
minutes.  We did not observe any correlation between the time it took
to answer questions and the participant background in data management
or even datalog.

{\em Group 2.} On Problem 1, the average score was 2.3, with the
following break-down: 6 participants received a perfect score, 3
received 2 out of 3, 1 had a score of 1, and 2 were not able to solve
the problem.  On Problem 2, 10 participants received a perfect score
and 1 got a score of 2 out of 3.  On Problem 3, 1 gave 5 good answers,
6 gave 3 good answers, 3 gave 2 good answers, and 2 gave no correct
answer.  The same two participants failed to answer Problems 1 and 3.

The test took longer for the participants without CS training.
Problem 1 took between 6 and 8 minutes to solve in this group, Problem
2 took between 5 and 8 minutes, and Problem 3 took between 4 and 12
minutes.

{\bf Remark.} We considered alternative ways in which a user can
interact with a \wdl system.  We are currently developing an interface
in which users will be able to write \wdl programs, but will also have
access to customizable canned queries implementing common
functionality. A SQL-based approach is not a natural choice, since
SQL does not accommodate distribution, which is central to \wdl.

{\bf In summary}, all technical and the majority of non-tech\-nical
participants of our study were able to both understand and write \wdl
programs correctly, with a minimal amount of training.  We observed a
difference between the technical and non-technical groups in terms of
both correctness and time to solution.  Two members of the
non-technical group were able to understand \wdl programs but were not
able to write programs on their own.  We believe that this issue will
be alleviated once an appropriate user interface becomes available.


\section{The WebdamLog System}
\label{sec:system}

In this section, we describe the architecture of the {\wdl} system.
We describe the implementation of the system, stressing the novel
features compared to standard datalog engines. \eat{ Performance
optimization techniques in support of very dynamic applications are
considered in Section~\ref{sec:dynamic}.}

\subsection{System architecture}
\label{sec:system:arch}

\begin{figure}
    \includegraphics[width=\columnwidth]{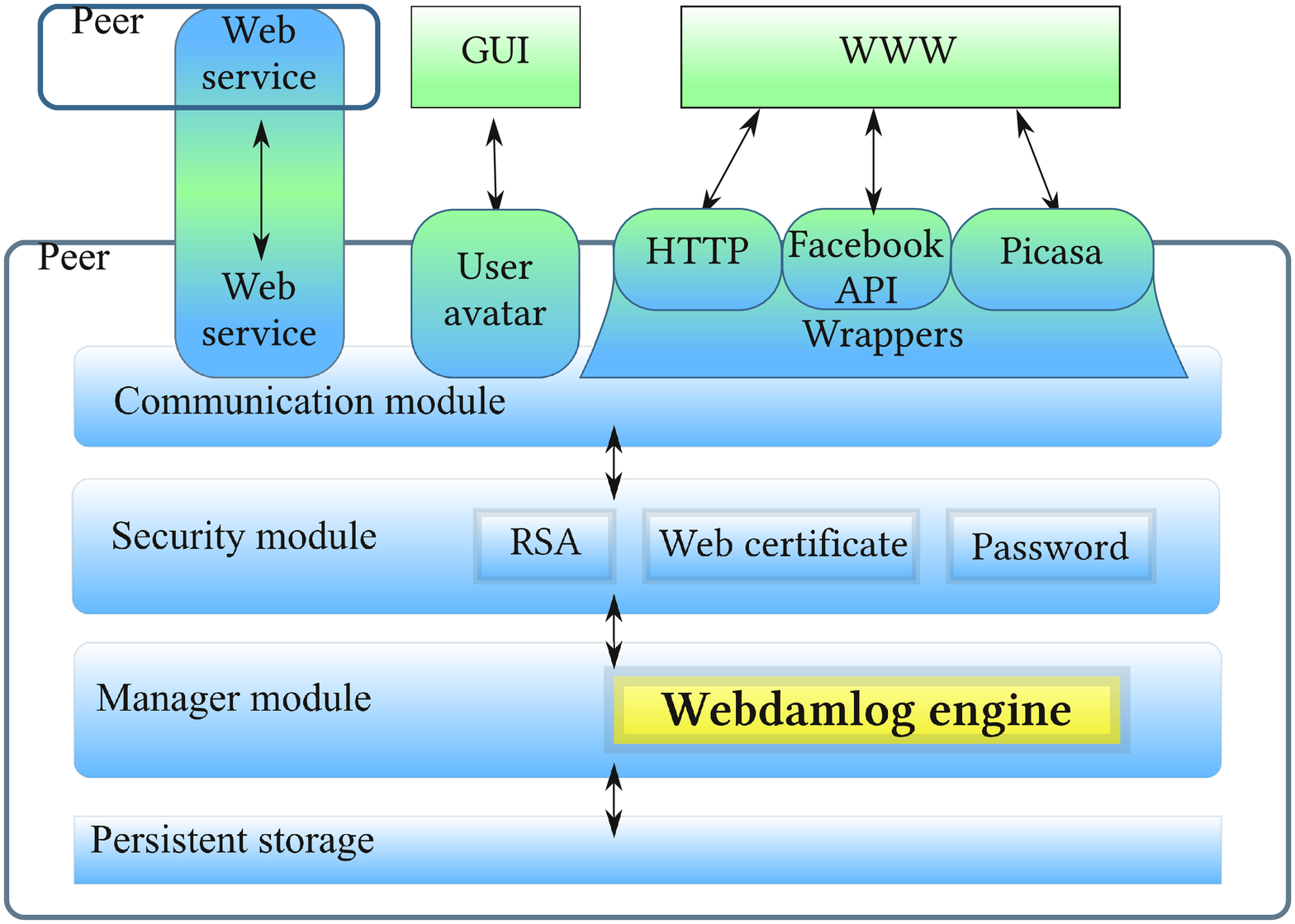}
    \caption{\wdl engine in a full \wdm peer}
    \label{fig:architecture_WE}
\end{figure}

Figure~\ref{fig:architecture_WE} shows the architecture of a \wdl
peer. Facts and rules are stored in a persistent store. The \wdl engine,
described in greater detail in the remainder of this section, retrieves
these facts and rules to process updates and answer queries coming from
the top layers. The Security module provides facilities for standard
access control mechanisms such as encryption, signatures and other
authentication protocols. The Communication module is responsible for
exchanging facts and rules with other peers.

Datalog evaluation has been intensively studied, and several
open-source implementations are available. We chose not to implement
yet another datalog engine, but instead to extend an existing one.  In
particular, we considered two open-source systems that are currently
being supported, namely, \bud~\cite{bud:website} from Berkeley
University and \textit{IRIS}~\cite{IRIS:website} from Innsbruck
University. The \textit{IRIS} system is implemented in Java\eat{, the
  language of \wde,} and supports the main strategies for efficient
evaluation of standard local datalog. The \bud system is implemented
in the Ruby scripting language, and initially seemed less promising
from a performance viewpoint.  However, \bud provides mechanisms for
asynchronous communication between peers, an essential feature for
\wdl.  In absence of a real performance comparison, the choice was not
easy. We finally decided in favor of \bud, both because of its support
for asynchronous communication, and because its scalability has been
demonstrated in real-life scenarios such as Internet routing.


\subsection{WebdamLog computation in Bud}
\label{sec:system:bud}

The \bud system supports a powerful datalog-like language introduced
in~\cite{bloom}.  Indeed, we see \bud (and use it) as a distributed
datalog engine with updates and asynchronous communications.

A \wdl computation consists semantically of a sequence of \emph{stages},
with each stage involving a single peer. Each stage of a \wdl peer
computation is in turn performed by a three-step \bud computation,
described next.  Note that we use the word \emph{stage} for \wdl and
\emph{step} for \bud:

\small
\begin{equation}
\begin{array}{r|ccc|ccc|l}
    \multirow{2}{*}{\dots} & \multicolumn{3}{|c|}{\mbox{Stage at peer p}} & \multicolumn{3}{|c|}{\mbox{Stage at peer q}} & \multirow{2}{*}{\dots} \\
    & \mbox{Step 1}   & \mbox{Step 2} &  \mbox{Step 3}  & \mbox{Step 1}   &  \mbox{Step 2} & \mbox{Step 3}   \\
\end{array}
\end{equation}
\normalsize

The 3 steps of a \wdl stage are as follows:
\begin{enumerate}

    \item Inputs are collected including input messages from other
      peers, clock interrupts and host language calls.

    \item Time is frozen; the union of the local store and of the
      batch of events received since the last stage is taken as an
      extensional database, and a \bud program is run to fixpoint.

    \item Outputs are obtained as side effects of the program,
      including output messages to other peers, updates to the local
      store, and host language callbacks.

\end{enumerate}

Observe that a fixpoint computation is performed at Step~2 by the
local datalog engine (namely the \bud engine). This computation is
based on a fixed program with no deletion over a fixed set of
extensional relations.  In Step~3, deletion messages may be produced,
along with updates to the set of rules and to the set of extensional
relations (for different reasons, which we will explain further). Note
that all this occurs \emph{outside} the datalog fixpoint computation.

Relations appearing in the rules are implemented as \bud
collections. \bud distinguishes between three kinds of key-value sets:
\begin{enumerate}
    \item A \emph{table} keeps a fact until an explicit delete order
      is received. We use tables to support \wdl extensional relations.

      \item A \emph{scratch} is used for storing results of intermediate
      computation. We use \emph{scratch} collections to implement \wdl
      local intentional relations. It is emptied at Step~1 and receives
      facts during fixpoint computation at Step~2.

    \item A \emph{channel} provides support for asynchronous
      communications. It records facts that have to be sent to other
      peers. We use channels for that and in particular for messages
      related to installing or removing delegations.
\end{enumerate}

As in \wdl, facts in a peer are consumed by the engine at each firing
of the peer (each stage).  To make facts persistent, they have to be
re-derived by the peer at each stage.  This is captured in our
implementation by assuming that rules re-derive extensional facts
implicitly, unless a deletion message has been received.

We observe a subtle point that lead us to not fully adopt the original
semantics of \wdl, as described in~\cite{webdamlog}.  There, we
assumed for simplicity that messages are transmitted instantaneously.
This assumption is not realistic in practice, and does not hold in our
implementation.  Since communications are asynchronous, there is no
guarantee in \wdl as to when a fact written to a channel will be
received by a remote peer.

\eat{
We observe two subtleties that lead us to not fully adopt the original
semantics of \wdl.

\begin{enumerate}

\item Since communications are asynchronous, there is no guarantee in
  \wdl as to when a fact written to a channel will be received by a
  remote peer. This is a departure from the original semantics of
  \wdl, which considered, for simplicity, that messages are
  transmitted instantaneously.  We depart from the original semantics
  because it imposes some form of synchronization, which may limit
  performance.

\item A subtlety is that deletions local to a peer are sent to the
  peer itself using a channel.  In the current system, there is
  therefore no guarantee that local deletion messages are received by
  the peer before the next stage is started. This is also a slight
  departure from the original semantics of \wdl that we do not see as
  important. If desired, this could be fixed.

\end{enumerate}
}

\subsection{Implementing WebdamLog rules}
\label{sec:system:wdl}

We now describe how \wdl rules are implemented on top of \bud.  We
distinguish between 4 cases.  This brings us to revisit the semantics of
\wdl (from Section~\ref{sec:wdl}) with a focus on implementation. As in
Section~\ref{sec:wdl}, whether a rule in a peer \peer{p} is \emph{local}
(i.e., all relations occurring in the rule body are \peer{p}-relations)
plays an important role. We consider 4 cases.  The last case (Case F)
focuses on the use of variables for relation and peer names. For the
first 3 cases, we ignore such variables.

\medskip\textbf{A-B-C. Simple local rules.} In this cases, i.e., local
rules with either an extensional relation or a local intentional
relation in the head, \wdl rules can be directly supported by identical
\bud rules.  (This takes care of local deduction as in datalog (A),
messages for local updates (B) and messages to other peers (C).)

\medskip\textbf{D. Local with non-local intentional head.} From an
implementation viewpoint, this case is more tricky. We illustrate it
with an example. Consider an intentional relation \relduo{s_0}{q}
defined in the distributed setting by the following two rules:
\begin{tabbing}
    [at p1] \quad \reltri{s_0}{q}{X,Y} :-
    \reltri{r_1}{p_1}{X,Y}
    \\
    {[}at p2] \quad \reltri{s_0}{q}{X,Y} :-
    \reltri{r_1}{p_2}{X,Y}
\end{tabbing}
Intuitively, the two rules specify a view relation \relduo{s_0}{q} at
\peer{q} that is the union of two relations \relduo{r_1}{p_1} and
\relduo{r_1}{p_2} from peers \peer{p_1} and \peer{p_2},
respectively. Consider a possible naive implementation that would
consist in materializing relation $s_0$ at \peer{q}, and having
\peer{p_1} and \peer{p_2} send update messages to \peer{q}.  Now suppose
that a tuple $\langle 0,1 \rangle$ is in both \relduo{r_1}{p_1} and
\relduo{r_1}{p_2}. Then it is correctly in \relduo{s_0}{q}. Now suppose
that this tuple is deleted from \relduo{r_1}{p_1}. Then a deletion
message is sent to \peer{q}, resulting in wrongly deleting the fact from
\relduo{s_0}{q}.

The problem arises because the tuple $\langle 0,1 \rangle$ originally
had two reasons to be in $s_0$, and only one of the reasons
disappeared. To avoid this problem, we could use the \emph{provenance}
of the fact $\langle 0,1 \rangle$ in \relduo{s_0}{q}.

A general approach for tracking provenance in our setting, and to
using it as basis for performance optimizations, is part of ongoing
work, and is outlined in Section~\ref{sec:prov_abbrev}.  For now, we
can implement the following \bud rules at \peer{p_1}, \peer{p_2} to
correctly support the two rules:

\begin{tabbing}
    [at p1] \quad \reltri{s_{0p1}}{q}{X,Y}:- \reltri{r_1}{p_1}{X,Y}\\
    {[}at p2] \quad \reltri{s_{0p2}}{q}{X,Y}:- \reltri{r_1}{p_2}{X,Y}\\
    {[}at q] \quad \reltri{s_0}{q}{X,Y}:- \reltri{s_{0p1}}{q}{X,Y}\\
    {[}at q] \quad \reltri{s_0}{q}{X,Y}:- \reltri{s_{0p2}}{q}{X,Y}
\end{tabbing}

\eat{To be precise, we have to explain the nature of the newly created
relations \relName{s_{0p1}} and \relName{s_{0p2}}. In fact, they may be
either intentional, in which case the view has to be computed on demand,
or extensional, in which case the view is materialized.}  Note that
relations \relName{s_{0p1}} and \relName{s_{0p2}} may be either
intentional, in which case the view is computed on demand, or
extensional, in which case the view is materialized.

\medskip\textbf{E. Non-local rules.} We consider non-local rules with
extensional head.  (Non-local rules with intentional head are treated
similarly.)  An example of such a rule is:
\begin{tabbing}
    [at p] \quad \reltri{r_0}{q}{\tuple{X_0}}:-
    \reltri{r_1}{q_1}{\tuple{X_1}},\dots,\reltri{r_i}{q_i}{\tuple{X_i}},\dots
\end{tabbing}
with \peer{q_1}$= \ldots =$ \peer{q_{i-1}} = \peer{p}, \peer{q_i} =
\peer{q} $\not =$ \peer{p}, and with each \tuple{X_j} denoting a tuple
of terms. If we consider atoms in the body from left to right, we can
process at \peer{p} the rule until we reach
$\reltri{r_i}{q}{\tuple{X_i}}$.  Peer \peer{p} does not know how to
evaluate this atom, but it knows that the atom is in the realm of
\peer{q}.  Therefore, peer \peer{p} rewrites the rule into two rules, as
specified by the formal definition of delegation in \wdl:

\begin{tabbing} [at p] \reltri{mid}{q}{\tuple{X_{mid}}} :-
    \reltri{r_1}{p}{\tuple{X_1}},\dots,\reltri{r_{i-1}}{p}{\tuple{X_{i-1}}}
    \\
    {[}at q] \reltri{r_0}{q}{\tuple{X_0}} :- \reltri{mid}{q}{\tuple{X_{mid}}}, \reltri{r_{i}}{q}{\tuple{X_{i}}},\dots
\end{tabbing}
where \emph{mid} identifies the message, and notably encodes, (i) the
identifier of the original rule, (ii) that the rule was delegated by
\peer{p} to \peer{q}, and (iii) the split position in the original rule.
The tuple \tuple{X_{mid}} includes the variables that are needed for the
evaluation of the second part of the rule, or for the head.  Observe
that the first rule (at \peer{p}) is now local. If the second rule,
installed at \peer{q}, is also local, no further rewriting is needed.
Otherwise, a new rewriting happens, again splitting the rule at
\peer{q}, delegating the second part of the rule as appropriate, and so
on.

Observe that an evolution of the state of \peer{p} may result in
installing new rules at \peer{q}, or in removing some delegations.
Deletion of a delegation is simply captured by updating the predicate
guarding the rule.  Insertion of a new delegation modifies the program
at \peer{q}.  Note that in \bud the program of a peer is fixed, and so
adding and removing delegations is a novel feature in \wdl.
Implementing this feature requires us to modify the \bud program of a
peer. This happens during Step 1 of the \wdl stage.

\medskip\textbf{F. Relation and peer variables.}  Finally, we consider
relation and peer variables.  In all cases presented so far, \wdl rules
could be compiled statically into \bud rules.  This is no longer
possible in this last case.  To see this, consider an atom in the body
of a rule. Observe that, if the peer name in this atom is a variable,
then the system cannot tell before the variable is instantiated whether
the rule is local or not. Also, observe that, if the relation name in
this atom is a variable, then the system cannot know whether that
relation already exists or not.  In general, we cannot compile a \wdl
rule into \bud until all peer and relation variables are instantiated.

To illustrate this situation more precisely, consider a rule of the
form:
\begin{tabbing}
    \reltri{r_0}{p}{\tuple{X_0}}:- \= \reltri{r_1}{p}{\var{X}},
    \dots,\reltri{\var{X}}{p}{\tuple{X_i}},\dots,
\end{tabbing}
where \relName{r_0}@\peer{p} is extensional and \var{X} is a variable.
This particular rule is relatively simple since, no matter how the
variable is instantiated, the rule falls into the simple case
\textbf{B}.  However, it is not a \bud rule because of the variable
relation name \var{X}.

Note that \wdl rules are evaluated from left to right, and a
constraint is that each relation and peer variable must be bound in a
previous atom. (This constraint is imposed by the language.)
Therefore, when we reach the atom \reltri{\var{X}}{p}{\tuple{X_i}},
the variable \var{X} has been instantiated.

To evaluate this rule, we use two \wdl stages of the peer. In the first
stage, we bind \var{X} with values found by instantiating
$\reltri{r_1}{p}{\var{X}}$. Suppose that we find two values for \var{X},
say {\em t$_1$} and {\em t$_2$}. We always wait for the next stage to
introduce new rules (there are two new rules in this case). More
precisely, new rules are introduced during Step 1 of the \wdl
computation of the next stage. In the example, the following rules are
added to the \bud program at $p$:
\begin{tabbing}
    \reltri{r_0}{p}{\tuple{X_0}}:- \reltri{t1}{p}{\tuple{X_i}},\dots,
    \\
    \reltri{r_0}{p}{\tuple{X_0}}:- \reltri{t2}{p}{\tuple{X_i}},\dots,
\end{tabbing}
Observe that, even in the absence of delegation, having variable
relation and peer names allows the \wdl engine to produce new rules at
run time, possibly leading to the creation of new relations.  This is a
distinguishing feature of our approach, and is novel to \wdl and to our
implementation.  \eat{because of these variables in relation or peer
names, the \wdl engine produces rules at run-time unlike a classic
datalog engine or even the \bud engine.  This could also possibly lead
to creating new relation names.}

This example uses a relation name variable. Peer name variables are
treated similarly.  Observe that having a peer name variable, and
instantiating it to thousands of peer names, allows us delegating a rule
to thousands of peers. This makes distributing computation very easy
from the point of view of the user, but also underscores the need for
powerful security mechanisms. Developing such mechanisms is in our
immediate plans for future work.

\subsection{Running the fixpoint}
\label{sec:adding-facts-rules}

The \bud engine evaluates the fixpoint using the semi-naive
algorithm\eat{~\cite{Alice13}}, i.e., \bud saturates one stratum after
another according to a stratification given by the {\em dependency
  graph}. The dependency graph is a directed hyper-graph with
relations as nodes, and with a hyper-edge from relations $s_i$ to
relation $r$ if there is a rule in which all $s_i$ appear in the body
and $r$ appears in the head.  Since this is classic material, we omit
the details but observe that, since \wdl rules may be added or removed
at run-time, the program evolves, leading to changes in the dependency
graph.  Therefore, the dependency graph is recomputed at step~1 of a
\wdl stage when receiving new rules, and remains fixed for the
following step~2. The \wdl engine pushes further the differentiation
technique that serves as basis of the semi-naive algorithm.

Although, according to \wdl semantics, facts are consumed and possibly
re-derived, it would be inefficient to recompute the proof of
existence of all facts at each stage. Between two consecutive stages,
each relation keeps a cache of its previous contents.  This cache may
be invalidated by \wdl if a newly installed rule creates a new
dependency for this relation. Note that \bud already performs cache
invalidation propagation for facts, which we adapt to fit \wdl
semantics. This incremental optimization across stages allows us to
run the fixpoint computation only on the relations that may have
changed since the previous stage.

\subsection{Maintaining dynamic peer state}
\label{sec:prov_abbrev}

A \wdl system executes in a highly dynamic environment, where peer
state frequently changes, in terms of both data and program, and where
peers may come and go.  This is a strong departure from datalog-based
systems such as \bud that assume the set of peers and rules to be
fixed.  As part of our ongoing work, we are focusing on efficiently
supporting dynamic changes in peer state, with the help of a novel
kind of a \emph{provenance graph}.

We use provenance graphs to record the derivations of \wdl facts and
rules, and to capture fine-grained dependencies between facts, rules,
and peers.  We build on the formalism proposed
in~\cite{provenance:semiring}, where each tuple in the database is
annotated with an element of a provenance semiring, and annotations
are propagated through query evaluation.  Provenance can be used for a
number of purposes such as explaining query results or system
behavior, and for debugging.  Our primary use of provenance is to
optimize performance of \wdl evaluation in presence of deletions.  We
are also currently investigating the use of provenance for enforcing
{\em access control} and for detecting access control violations.


\section{Experimental Evaluation}
\label{sec:experiments}

The goal of the experimental evaluation is to verify that \wdl
programs can be executed efficiently.  We show here that rewriting and
delegation can be implemented efficiently. \eat{ In
  Section~\ref{sec:exp:dynamism}, we demonstrate that the data and
  programs of rapidly evolving peers can be efficiently maintained
  using provenance graphs.}

In the experiments, we used synthetically generated data.  All
experiments were conducted on up to 50 Amazon EC2 micro instances,
with 2 \wdl peers per instance.  Micro-instances are virtual machines
with two process units, Intel(R) Xeon(R) CPU E5507 @2.27GHz with 613
MB of RAM, running Ubuntu server 12.04 (64-bit).  All experiments were
executed 4 times with a warm start. We report averages over 4
executions.

\eat{
The examples are inspired by an implementation (in a slightly
simplified form) of the motivating example described in
Section~\ref{sec:exa}, in which friends of Alice and Bob are making a
photo album for them as a wedding present.  This example is
representative of a number of real tasks where many peers collaborate
by sharing information. The experiments are designed to capture the
salient features of such applications.

\serge{REMOVE: The only simplification for the purpose of the
experiments is that we assume, to simplify, that each friend keeps
their photos on their peer.
As in the example in Section~\ref{sec:exa}, we work with 3 designated
peers representing Alice, Bob and Sue, and with a varying number of
peers representing friends of Alice and Bob. Peers \peer{alice} and
\peer{bob} each contain an extensional relation
\relName{friends}(\var{name}). The number of facts in these relations
allows controlling the degree of distribution.  Each peer representing
a friend of Alice or Bob contains two extensional relations:
\relName{photos}(\var{photoId}) and
\relName{features}(\var{photoId},\var{tag}), storing, respectively,
the ids of photos and the tags describing the contents of the photos.}
}


{\bf The cost of delegation.} We now focus is on measuring \wdl
overhead in dealing with delegations. Recall the \bud steps performed
by each peer at each \wdl stage, described in
Section~\ref{sec:system:bud}.  We can break down each step into
\wdl-specific and \bud-specific tasks as follows:

\begin{enumerate}\addtolength{\itemsep}{-0.5\baselineskip}
    \item Inputs are collected
    \begin{enumerate}
       \item \textbf{\bud} reads the input from the network and
         populates its channels.\label{breakdown:1}
       \item \textbf{\wdl} parses the input in channels and updates
         the dependency graph with new rules. The dependency graph is
         used to control the rules that are used in the semi-naive
         evaluation (see
         Section~\ref{sec:adding-facts-rules}). \label{breakdown:2}
    \end{enumerate}
    \item Time is frozen
    \begin{enumerate}
        \item \textbf{\bud} invalidates each $\Delta$ (used by the
          semi-naive evaluation) that has to be reevaluated because it
          corresponds to a relation that may have
          changed. \label{breakdown:3}
        \item \textbf{\wdl} invalidates $\Delta$ according to program
          updates. Moreover, \wdl propagates deletions. (Recall that
          the semi-naive evaluation deals only with tuple
          additions.) \label{breakdown:4}
        \item \textbf{\bud} performs semi-naive fixpoint evaluation for
        all invalidated relations, taking the last $\Delta$ for
        differentiation.\label{breakdown:5}
    \end{enumerate}
  \item Outputs are obtained
    \begin{enumerate}
        \item \textbf{\wdl} builds packets of rules and updates to
          send.\label{breakdown:6}
        \item \textbf{\bud} sends packets.\label{breakdown:7}
    \end{enumerate}
\end{enumerate}

We report the running time of \wdl as the sum of
Steps~\ref{breakdown:2}, \ref{breakdown:4} and~\ref{breakdown:6}, and
the running time of \bud as the sum of Steps~\ref{breakdown:1},
\ref{breakdown:3}, \ref{breakdown:5} and \ref{breakdown:7}. All
running times are expressed in percentage of the total running time,
which is measured in seconds. For each experiment, we will see that
the running time of \wdl-specific phases is reasonable compared to the
overall running time.

\eat{
For the experiments in this section, we use \wdl rules involving {\em
  only extensional relations}, both in the head and in the body.  We
also support rules with intentional relations in the head and in the
body. But for such rules, an essential optimization consists in
deriving \emph{only the relevant data and delegated rules}.  We intend
to conduct experiments with such rules when our system supports
optimizations in the style of Magic Set.  }

{\bf Non-local rules.} In the first experiment, we evaluate the
running time of a non-local rule with an extensional head.  Rules of
this kind lead to delegations.  We use the following rule:

\begin{verbatim}
[at alice]
join@sue($Z) :- rel1@alice($X,$Y), rel2@bob($Y,$Z)
\end{verbatim}
This rule computes the join of two relations at distinct peers
(\relName{rel1}@\peer{alice} and \relName{rel2}@\peer{bob}), and then
installs the result, projected on the last column, at the third peer
(\relName{join}@\peer{sue}). Relations \relName{rel1}@\peer{alice} and
\relName{rel2}@\peer{bob} each contain 1~000 tuples that are pairs of
integers, with values drawn uniformly at random from the 1 to 100
range.  In the next table, we report the total running time of the
program at each peer, as well as the break-down of the time into \bud
and \wdl.


\begin{center}
    \begin{tabular}{l|c|c|r|}
        \cline{2-4}
        & \wdl & \bud  & total \\ \cline{1-4}
        \multicolumn{1}{|c|}{\peer{alice}} & 10.8\% & 89.2\% & 0.10s \\ \cline{1-4}
        \multicolumn{1}{|c|}{\peer{bob}} & 4.0\% & 96.0\% & 0.87s \\ \cline{1-4}
        \multicolumn{1}{|c|}{\peer{sue}} & 0.7\% & 99.3\% & 0.02s \\ \cline{1-4}
    \end{tabular}
\end{center}


The portion of the overall time spent on \wdl computation on
\peer{alice} is fairly high:~10.8\%.  This is because that peer's work
is essentially to delegate the join to \peer{bob}.  Peer \peer{bob}
spends most of its time computing the join, a \bud computation.  Peer
\peer{sue} has little to do.  As can be seen from these numbers, the
overhead of delegation is small.

{\bf Relation and peer variables.} In the second experiment, we
evaluate the execution time of a \wdl program for the distributed
computation of a union.  The following rule uses relation and peer
variables and executes at peer \peer{sue}:
\begin{verbatim}
[at sue]
union@sue($X) :- peers@sue($Y,$Z), $Y@$Z($X)
\end{verbatim}

The relation \relName{peers}@\peer{sue} contains 12 tuples
corresponding to 3 peers (including \peer{sue}) with 4 relations per
peer. Thus, the rule specifies a union of 12 relations. Each relation
participating in the union contains 1~000 tuples, each with a single
integer column, and with values for the attribute drawn independently
at random between 1 and 10~000.


\begin{center}
    \begin{tabular}{l|c|c|r|}
        \cline{2-4}
        & \wdl & \bud & total \\ \cline{1-4}
        \multicolumn{1}{|c|}{\peer{sue}} & 9.9\% & 90.1\% & 1.04s \\ \cline{1-4}
        \multicolumn{1}{|c|}{\peer{remote1}} & 1.1\% & 98.9\% & 0.04s \\ \cline{1-4}
        \multicolumn{1}{|c|}{\peer{remote2}} & 1.3\% & 98.7\% & 0.04s \\ \cline{1-4}
    \end{tabular}
\end{center}


Observe that \peer{sue} does most of the work, both delegating rules
and also computing the union. The \wdl overhead is 9.9\%, which is
still reasonable. The running time on remote peers is very small, and
the \wdl portion of the computation is negligible.

{\bf QSQ-style optimization.} In this experiment, we measure the
effectiveness of an optimization that can be viewed as a distributed
version of query subquery (QSQ)~\cite{Vieille86-QSQ}, where only the
relevant data are communicated at query time.  More precisely, we
consider the following view \relName{union2} on peer \peer{sue},
defined as the union of two relations.
\begin{verbatim}
[at sue]
union2@sue($name,$X) :- friendPhotos@alice($name,$X)
union2@sue($name,$X) :- friendPhotos@bob($name,$X)
\end{verbatim}
Suppose we want to obtain the photos of Charlie, i.e. the tuples in
\relName{union2} that have the value ``Charlie'' for first attribute.
We vary the number of facts in \relName{friendPhotos}@\peer{alice} and
\relName{friendPhoto}@\peer{bob} that match the query.  We compare the
cost of materializing the entire view to answer the query to that of
installing only the necessary delegations computed at query time to
compute the answer.

\begin{figure}[tb!]
    \centering
    \includegraphics[width=\columnwidth]{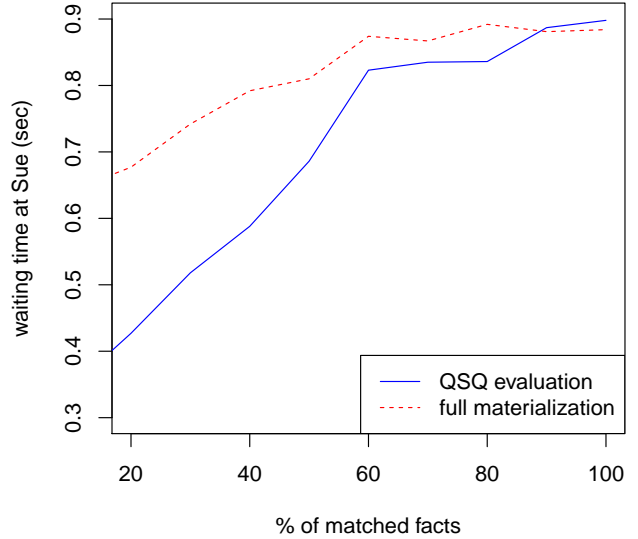}
    \caption{Distributed QSQ optimization}
    \label{fig:matviewVSdelegation}
\end{figure}

Results of this experiment are presented in
Figure~\ref{fig:matviewVSdelegation}.  We report the waiting time at
\peer{sue}.  As expected, QSQ-style optimization brings important
performance improvements (except when almost all facts are selected).
This shows its usefulness in such a distributed setting.

\section{Related work}
\label{sec:related}


The \wdl language is motivated by previous work on the
\emph{WebdamExchange} system~\cite{webdamexchange:paper}.  The system
described there could automatically adapt to a variety of protocols
and access methods found on the Web, notably for localizing data and
for access control~\cite{webdamexchange:demo}.  In developing toy
applications with \emph{WebdamExchange}~\cite{webdamexchange:demo}, we
realized the need for a logic that could be used (i) to
\emph{declaratively} specify applications and (ii) to exchange
application logic between peers.  This motivated the introduction of
\wdl~\cite{webdamlog}, a language based on rules that can run locally
and be exchanged between peers.

Distributed data management has been studied since the earliest days
of databases~\cite{OV11}. The fact that it is possible to access data
from several data sources has been studied under various names,
notably multi-databases or federated databases. The setting we
consider is in the spirit of \emph{peer-to-peer databases} with
autonomous and heterogeneous data sources. Of course, standard query
optimization techniques developed for distributed database systems are
relevant here. We insist in particular on the techniques that are more
relevant to our setting, which is based on datalog. One should mention
that there have been a number of works on parallel or distributed
evaluation of datalog, e.g.,~\cite{dQSQ,HAC90}.

The use of declarative languages, in particular datalog extensions,
for distributed data management has already been advocated, e.g.,
in~\cite{activexml-vldb,AbiteboulP07}.  There has recently been
renewed interest in this approach~\cite{Hellerstein10}. Several
systems have been developed based on the declarative
paradigm~\cite{NetLog, Overlog, NDLog}, with performance comparable to
that of systems based on imperative languages.  Our implementation
uses the \bud system~\cite{bud:website}.  The language
Dedalus~\cite{dedalus} has been proposed as a formal foundation for
\bud.  We prefer here to use the language \wdl, in particular because
it features delegation.

Most classic optimization techniques for datalog are relevant to our
work, in particular, semi-naive evaluation that is supported by \bud.
We also considered the query-subquery optimization~\cite{Vieille86-QSQ}
as adapted to the distributed context in~\cite{dQSQ}.

\eat{
Distributed datalog processing brings about new performance
challenges, particularly when data and programs evolve very rapidly.
We use provenance in support of efficient \wdl reasoning, building on
recent works.  A data model for provenance applicable to datalog was
first formalized in~\cite{provenance:semiring}, and a provenance query
language was proposed in~\cite{proQL}.  Provenance has been used in
distributed systems to propagate updates according to schema mappings
and trust~\cite{UpdateMappingProv,orchestra}.  However, unlike in our
system, provenance there is maintained centrally.  With respect to
using provenance for the evaluation of distributed datalog, the ExSPAN
system~\cite{exspan} is most closely related to our work, because they
also maintain provenance in a distributed manner.  An important
difference between ExSPAN and our system is that our rules may carry
provenance, and so are represented as nodes in the provenance graph.
Inspired by recent work on fine-grained workflow
provenance~\cite{lipstick}, we augment the provenance graph by peer
labels, in support of further performance optimizations.
}


\section{Conclusion}
\label{sec:conc}


This paper presents an implementation of the \wdl language, introduced
in~\cite{webdamlog}.  The two main challenges for such an approach are
(i) the difficulty of writing rules for non-technical users and (ii)
the difficulty to offer good performance:
\begin{itemize}

\item With respect to (i), we present a user study that very
  promisingly shows that the participants (many of them not computer
  scientists) are able to understand and write simple rules.

\item With respect to (ii), we benefit from previous datalog
  optimization techniques and efficient network communication by
  relying on the \bud system to support the basic functionality of
  distributed datalog. We show that the higher level features of \wdl,
  notably delegation, can be supported efficiently using logical rule
  rewriting.

\end{itemize}
All this demonstrates the feasibility of an approach based on \wdl to
support exchanges of data and rules between rapidly evolving peers in
a distributed and dynamic environment.

In the future, we are considering the following directions:
\begin{description}

\item [Access control] One of the bases of \wdl is that a peer can
  locally install rules that are specified by another peer. Clearly,
  this is potentially very risky. Access control is therefore of
  paramount importance. We plan to work on access control, and in
  particular investigate the use of provenance for enforcing access
  control and for detecting access control violations.

 \item [Interface] Our user study demonstrated that \wdl is
   appropriate for specifying distributed data management tasks.  We
   are in the process of developing a user interface for the \wdl
   system. We also plan to conduct a follow-up user study (i) drawing
   from a larger pool of participants, (ii) including more
   participants without any CS training, and (iii) testing the
   usability of other aspects of the language, notably intentional
   vs. extensional predicates.

 \item [Application] We intend to demonstrate the use of our system
   with complete applications, e.g., for social networks and personal
   data management. 

 \item [Optimization] We are currently developing a provenance-based
   approach for efficiently supporting changes in program state.
   Also, an optimization technique based on map-reduce and intense
   parallelism has been proposed for
   datalog~\cite{AfratiMapRedRecQuery}.  It would be interesting to
   consider such an approach in our distributed setting.

\end{description}


\vspace{2em}
\bibliographystyle{abbrv}
\bibliography{biblio}
\vspace{2em}
\balance

\end{document}